\documentclass[epj]{svjour}

\usepackage{epsfig}

\usepackage{graphicx,latexsym,xcolor}
\usepackage{amsmath}

\global\long\def\bege{\begin{equation}}
\global\long\def\ende{\end{equation}}

\global\long\def\begal{\begin{align}}
\global\long\def\endal{\end{align}}

\title{Energetical self-organization of a few strongly interacting particles}

\author{Ioannis Kleftogiannis\inst{1} \and Ilias Amanatidis\inst{2}}

\institute{                    
  \inst{1} Physics Division, National Center for Theoretical Sciences, Hsinchu 30013, Taiwan\\
  \inst{2} Department of Physics, Ben-Gurion University of the Negev, Beer-Sheva 84105, Israel
}

\date{\today \\ \email{ph04917@yahoo.com, eliasamanatidis1@hotmail.com}}
\abstract{
We study the quantum self-organization of a few interacting particles with strong short-range interactions. The physical system is modeled via a 2D Hubbard square lattice model, with a nearest-neighbor interaction term of strength U and a second nearest-neighbor hopping t. For t=0 the energy of the system is determined by the number of bonds between particles that lie on adjacent sites in the Hubbard lattice. We find that this bond order persists for the ground and some of the excited states of the system, for strong interaction strength, at different fillings of the system. For our analysis we use the Euler characteristic of the network/graph grid structures formed by the particles in real space (Fock states), which helps to quantify the energetical(bond) ordering. We find multiple ground and excited states, with integer Euler numbers, whose values persist from the $t=0$ case, for strong interaction $U>>t$. The corresponding quantum phases for the ground state contain either density-wave-order(DWO) for low fillings, where the particles stay apart form each other, or clustering-order(CO) for high fillings, where the particles form various structures as they condense into clusters. In addition, we find various excited states containing superpositions of Fock states, whose probability amplitudes are self-tuned in a way that preserves the integer value of the Euler characteristic from the $t=0$ limit.
}

\begin{document}

\authorrunning{I.Kleftogiannis and Ilias Amanatidis} 
\titlerunning{Energetical self-organization of a few strongly interacting particles}
\maketitle

\section{Introduction}
Strong interactions in quantum systems with many self-organizing particles, can give rise to phases of matter with many unusual and interesting properties.
For example extensive studies have been performed for quantum many-body systems via Hubbard models with various types of interactions between the particles, leading to diverse clustering phenomena and emergent many-body phases such as the density-wave, Mott-insulating and superfluid phases \cite{cdw1,cdw2,cdw3,cdw4,masella,2d,excited,slava,2017,fqhe1d}. 
Other examples are those related to quantum correlations (entanglement) and topology which have been dubbed topological orders\cite{spinhaldane,haldane0,AKLT,Levin,Gu,kitaev2,kitaev3,alba,alba1,hen,Hamma,Calabrese,Pollmann,amico,horodecki}. In these cases the physical system undergoes phase transitions, that do not obey Landau's symmetry breaking mechanism that describes the more common phase transitions, that occur also in classical systems.
Topology and quantum correlations are strongly tied
in topologically ordered phases of matter, giving
rise to massively entangled states of matter
with unconventional features such as fractionally quantized excitations and non-standard particle statistics, like anyonic braiding statistics\cite{Tsui,Laughlin,Stormer,Li,haldane_geometry}. Topological numbers can be used to categorize different topological phases of matter, such as the Chern number, the winding number or the topological entanglement entropy\cite{kitaev2,kitaev3,Li,haldane_geometry}.

In this paper we study the quantum self-organization
of interacting particles in two-dimensions(2D), modeled via minimal 2D Hubbard square lattice models, containing only a nearest-neighbor interaction between the particles and a second nearest-neighbor hopping. The model is motivated by our previous analysis of an 1D Hubbard chain with nearest-neighbor interaction and second nearest-neighbor hopping that gives rise to many-body states with topological properties, related to the fractional-Hall-effect (FQHE) \cite{fqhe1d}. We demonstrate that the 2D model gives rise to various many-body states with unconventional properties, formed via the clustering of the particles at various energies and fillings, for strong interaction strength. In order to characterize the structural properties of these states we use the Euler characteristic(number) of the network/graph grid structures formed by the particles as they self-organize at different energies and fillings. We find various ground and excited states whose Euler number takes integer values, persisting from the case when the hopping term is absent and the self-organization of the particles is fully determined by the interaction term only. We study the structural properties of these states for various fillings.

\section{Model}
For our calculations we use a minimal Hubbard square lattice model of spinless particles, where only one particle is allowed per site, described by the Hamiltonian 
\begin{equation}
\begin{aligned}
& H = H_U+H_t \\
& H_U = U \sum_{x=1}^{L_{x}}\sum_{y=1}^{L_{y}}(n_{x,y}n_{x+1,y} +  n_{x,y}n_{x,y+1}) \\
& H_t=t\sum_{x=1}^{L_{x}}\sum_{y=1}^{L_{y}}(c_{x+2,y}^{\dagger}c_{x,y} +  c_{x,y+2}^{\dagger}c_{x,y} + h.c. )\\
\end{aligned}
\label{eq1}
\end{equation}
where $c_{x,y}^{\dagger},c_{x,y}$ are the creation and annihilation operators for spinless particles
at site with coordinates x,y in the lattice, while
$n_{xy}=c_{x,y}^{\dagger}c_{x,y}$ is the number operator.
We consider $L_{x}(L_{y})$ number of sites
along the x(y) direction, giving the total number of sites in the system $L=L_{x} \times L_{y}$. Also we assume that the system terminates with hard-wall boundary conditions in both directions, which is obtained after removing the hopping and interaction terms with $x>L_y$ and
$y>L_y$ from Eq. \ref{eq1}. Finally, we consider N particles distributed among the L sites. The filling of the system is $f=\frac{N}{L_{x} \times L_{y}}=\frac{N}{L}$.
Eq. \ref{eq1} could describe hard-core bosons or spinless fermions.The calculations presented in this manuscript are for hard-core bosons which satisfy the commutation relation $[c_{i},c_{j}^{\dagger}] = (1-2n_{i})\delta_{ij}$.
Our analysis using Eq.  \ref{eq1} can be considered as an extension of our previous study in 1D where we have shown the emergence of topological states at various fillings due to either density-wave or clustering order\cite{fqhe1d}. 

The interaction term $H_U$ lifts the energy of the system by U when two particles occupy nearest-neighboring (adjacent) sites in the Hubbard lattice. The hopping term $H_t$ allows the particles to hop between second nearest-neighboring sites in the Hubbard lattice. This type of hopping is crucial for the appearance of the quantum phases that we observe. The N particles distribute among L sites giving many possible particle configurations whose number is determined by the binomial \begin{equation}
D(L,N)=\binom{L}{N}.
\label{fock_size}
\end{equation}
This number counts the number of the Fock states,
which act as the basis states for the Hamiltonian matrix
given by Eq. \ref{eq1}, whose size is given by Eq. \ref{fock_size}.
The square lattice system described by Eq. \ref{eq1} and the resulting particle structures emerging, can be considered as grid graphs, whose Euler characteristic can be defined as 
\begin{equation}
\chi=N-M,
\label{euler}
\end{equation}
where the N particles act as vertices in the graph and $M$ is the number of edges between these vertices/particles, formed when two particles occupy nearest-neighboring (adjacent) sites in the Hubbard lattice. When $t=0$ then the energy E of the system is fully determined by the number of edges between the particles via $E=M U=(N-\chi)U$. The definition Eq. \ref{euler} of the Euler allows us to describe the clustering of the interacting particles in a graph mathematical language. The various clustering structures(graphs) emerging contain features like a variable number of disconnected clusters and closed vertex lines, known as induced(chordless) cycles in graph theory, where no two vertices of the cycle are connected by an edge that does not itself belong to the cycle. In the rest of the paper we use the term cycles
to denote the induced cycles.

From this perspective it is also useful to define the curvature at each site i of the Hubbard lattice by using the standard notion for tree or grid graphs\cite{2d,excited,chen,oliver}
\begin{equation}
K_{x,y}=\langle n_{x,y} \rangle - \frac{\langle d_{x,y} \rangle}{2},
\label{eq_10}
\end{equation}
where $\langle n_{x,y} \rangle$ is the particle density at site with coordinates x,y of the Hubbard lattice and $\langle d_{x,y} \rangle$ is the number of its nearest-neighboring particles (edges), which can be written as
$\langle d_{x,y} \rangle=\langle n_{x,y} \rangle (\langle n_{x,y+1} \rangle +\langle n_{x,y-1}\rangle+\langle n_{x+1,y} \rangle +\langle n_{x-1,y}\rangle)$. 
The Euler characteristic can be calculated by summing
the curvature over all the sites of the Hubbard lattice
\begin{equation}
\chi=\sum_{x=1}^{L_x} \sum_{y=1}^{L_y} K_{x,y}
\label{eq_13}
\end{equation}
as in the Gauss-Bonnet theorem of differential geometry,
where a curvature is integrated over the surface of a manifold to get its Euler characteristic.
We note that Eq. \ref{eq_13} is essentially the average
Euler over the Fock states whose superposition forms the many-body wavefunction of the system at a specific energy, given by
\begin{equation}
\chi=\sum_{i=1}^{D(L,N)} \chi_i |\Psi_i|^2,
\label{eq_14}
\end{equation}
where $\chi_i$($\Psi_i$) is the Euler number(probability amplitude) for each Fock state i.
In the rest of the paper the term Euler characteristic ($\chi$) refers to its average value over all the Fock states, unless explicitly stated otherwise.

\begin{figure}
\begin{center}
\includegraphics[width=0.9\columnwidth,clip=true]{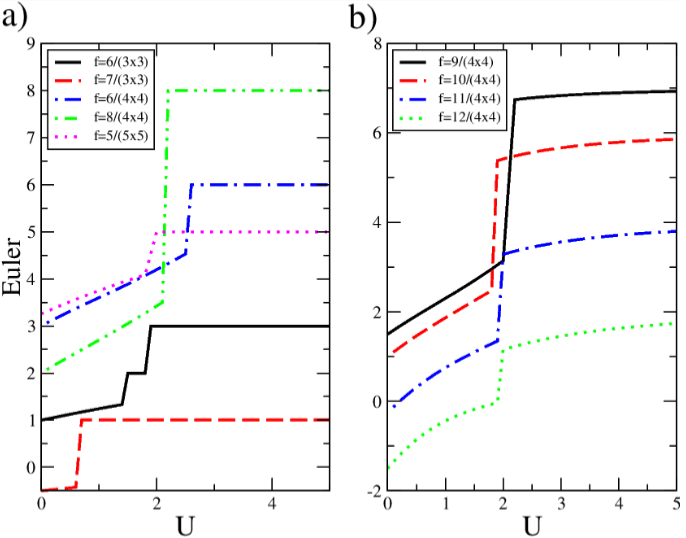}
\end{center}
\caption{The Euler characteristic $\chi$ for the ground state of system for fillings $f=\frac{N}{L_{x} \times L_{y}}$ with $L_x=L_y$, versus the strength of the interaction between the particles U. a)Cases where the system reaches quantum phases characterized by integer values of $\chi$ for large U. For $f\le\frac{1}{2}$ we have $\chi=N$ since the particles self-organize in density-wave-ordered(DWO) states without occupying adjacent sites in the Hubbard lattice. For $f>\frac{1}{2}$ we have $\chi<N$, whose integer values are determined by the number of edges in the clustering/graph structures formed by the interacting particles. b)Some respective cases for $f>\frac{1}{2}$ lacking the quantum phases, where the Euler characteristic approaches integer values asymptotically only, for $U \rightarrow \infty$.}
\label{fig1}
\end{figure}

\section{Ground State}
In figure \ref{fig1} we show the Euler characteristic $\chi$ versus the interaction strength U, for the ground state of square sample systems with $L_x=L_y$, corresponding to various fillings $f=\frac{N}{L_{x} \times L_{x}}$.
The cases shown in figure \ref{fig1}a reach quantum phases characterized by integer values of $\chi$ for sufficiently strong U, indicated by the flat steps(plateaus). The value of $\chi$ is determined by the empty space in the system i.e. the degree of spatial freedom of the interacting particles. For example the cases $f=\frac{6}{4 \times 4},\frac{8}{4 \times 4},\frac{5}{5 \times 5}$ all reach the value $\chi=N$. The corresponding states are linear superpositions of Fock states consisting of particle structures with density-wave-order(DWO), where all the particles stay apart from each other, not occupying adjacent sites in the lattice. The corresponding graph structures formed by the particles for each Fock state, lack any edges between the vertices (M=0) and therefore the average Euler number over all the Fock states, calculated via Eq. \ref{eq_13} or Eq. \ref{eq_14} is simply $\chi=N$.
We have observed a similar effect in the 1D version of the current model, consisting of Hubbard chains with a nearest-neighbor interaction term and a second-nearest neighbor hopping. The 1D DWO states lead to topological quantum phases for odd denominator fillings that correspond to odd number of sites in the Hubbard chains, leading to a realization of the fractional-quantum-Hall-effect (FQHE) in 1D\cite{fqhe1d}.

\begin{figure}
\begin{center}
\includegraphics[width=0.9\columnwidth,clip=true]{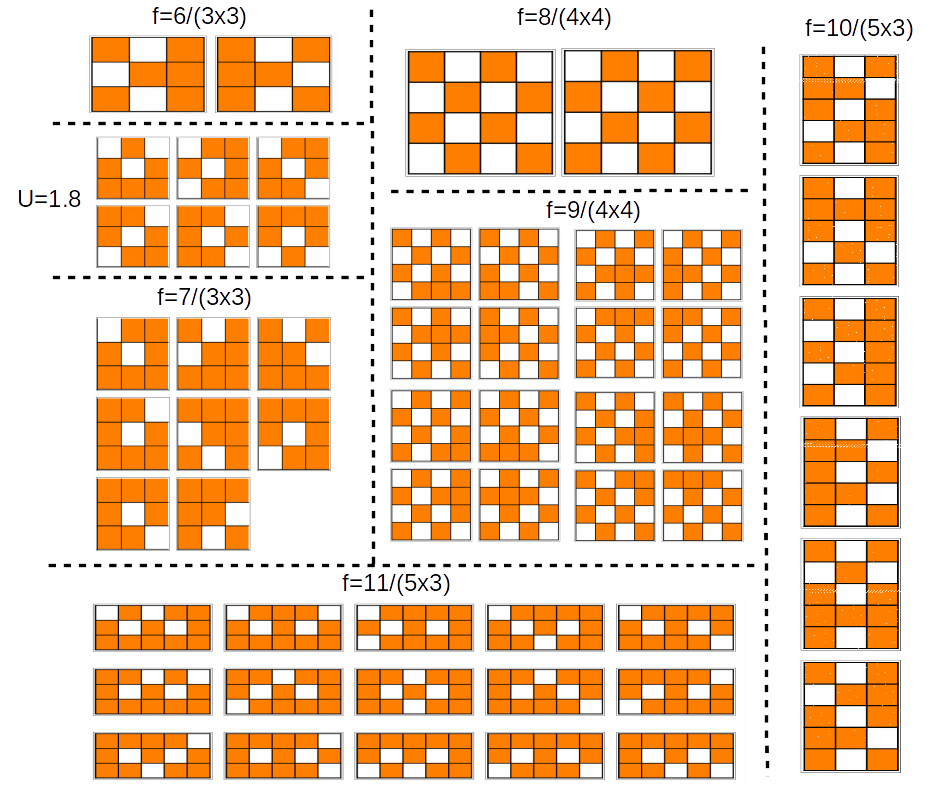}
\end{center}
\caption{Various particle structures (Fock states), whose superposition is the ground state of the system at various fillings with $t=1$ and $U=5$. Filled(empty) squares correspond to particles(holes) in the Hubbard lattice. The filling $f=\frac{8}{4 \times 4}$ contains two states with density-wave-order(DWO), where the particles don't simultaneously occupy adjacent sites in the Hubbard lattice, leading to a quantum phase with Euler $\chi=8$. The Fock states for the rest of the fillings shown, all contain edges between the graph structures formed by the particles, as they condense into clusters, leading to clustering-order(CO). The filling $f=\frac{7}{3 \times 3}$ contains two types of structures leading to the quantum phase with $\chi=1$. One structure consists of a single cluster containing all the particles with no closed vertex line structures (cycles)($\chi=1$). The other structure consists of a cluster with one cycle($\chi=0$) along with one single particle($\chi=1$). For $f=\frac{10}{5 \times 3}$ there are four clusters inside each Fock state arranged in various configurations. Each cluster contains one, four or seven particles, with no cycles. The Euler of each cluster is $\chi=1$, leading to the corresponding quantum phases with average Euler $\chi=4$. For $f=\frac{6}{3 \times 3}$ we show two cases for different U, the upper one $(U=5)$ with three clusters and the bottom one $(U=1.8)$ with two clusters, leading to respective quantum phases with $\chi=3$ and $\chi=2$, i.e. the number of clusters acts as a topological number, determining the Euler characteristic in this case. On the other hand, the states for fillings $f=\frac{9}{4 \times 4}$ and $f=\frac{11}{5 \times 3}$ do not lead to quantum phases characterized by integer Euler.}
\label{fig2}
\end{figure}

 In figure \ref{fig2} we show the Fock states for $\frac{8}{4 \times 4}$ containing the DWO, where the colored squares denote particles and the white squares denote empty sites in the Hubbard lattice. The system lies in a superposition of two states which contain two possible configurations of the particles where neighboring(adjacent) sites cannot be simultaneously occupied. 

On the other hand, the fillings $f=\frac{6}{3 \times 3},\frac{7}{3 \times 3}$ give Fock states where the particles cannot stay apart from each other, as there is not enough free space in the system and consequently they condense into clusters. For $f=\frac{6}{3 \times 3}$ the Euler number is $\chi=3$ (see figure \ref{fig1}), coinciding with the number of clusters in the Fock states, as can be seen in figure \ref{fig2}. The clusters consist of two single particles and a structure consisting of four adjacent particles with no closed vertex lines(cycles). All three clusters have $\chi=1$ resulting in total Euler $\chi=3$ for each of the two Fock states. A linear superposition of these two Fock states gives the quantum phase with $\chi=3$ shown in figure \ref{fig1}a for $f=\frac{6}{3 \times 3}$ at large U. Another interesting feature of $f=\frac{6}{3 \times 3}$ is an extra plateau at $\chi=2$ shown in figure \ref{fig1}a. As shown in figure \ref{fig2} this phase corresponds to particle structures with two clusters, containing one to five adjacent particles with no cycles. A case where the Euler number does not coincide with the number of clusters is the filling $f=\frac{7}{3 \times 3}$, which as shown in figure \ref{fig1}a, gives a quantum phase with $\chi=1$ for strong U. As shown in figure \ref{fig2} the corresponding Fock states contain a variable number of clusters, either a single cluster with all seven particles condensed, or two clusters consisting of a single particle with $\chi_{1}=1$ and a structure with one cycle with $\chi_{2}=0$, giving the total Euler, $\chi_{1}+\chi_{2}=1$. We notice that the different Fock states follow the same topology, since a line cluster is topologically equivalent to a single particle and the cluster with one cycle can always be shrunken down topologically to an empty site, so that the two Fock states for $f=\frac{7}{3 \times 3}$ can be smoothly deformed between each other. This property hints a topological character for the quantum phases characterized by the integer values of the Euler\cite{excited}.

In figure \ref{fig1}b we show several cases, all with an even total number of sites L and $f>\frac{1}{2}$, which lack the quantum phases characterized by integer values of $\chi$ for strong interaction strength U. Integer values of $\chi$ are reached only asymptotically for $U \rightarrow \infty$ for all fillings. Since there is not enough free space in the system, the particles cannot form DWO states, condensing instead into clusters. An example of the clustering structures for $f=\frac{9}{4 \times 4}$ is included in figure \ref{fig2}. The individual Euler number for each of these Fock states ranges from $\chi=5$ to $\chi=7$, as is the number of clusters.

\begin{figure}
\begin{center}
\includegraphics[width=0.9\columnwidth,clip=true]{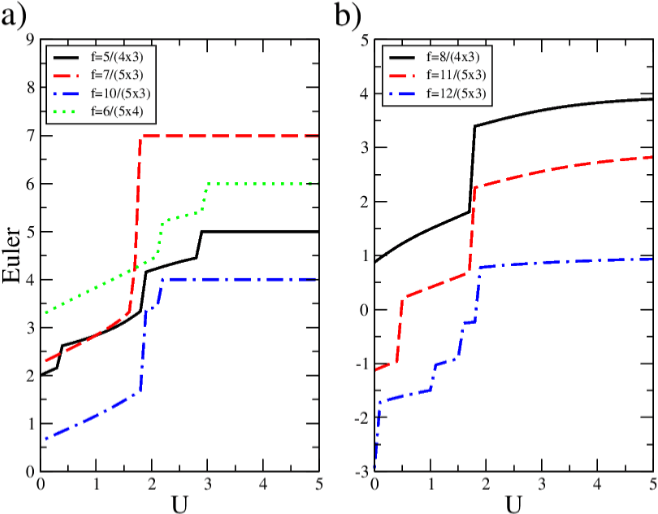}
\end{center}
\caption{The Euler characteristic $\chi$ for non-square sample systems. a)Cases where quantum phases are reached corresponding
to integer $\chi$. b)Several cases lacking the quantum phases, where integer $\chi$ is reached asymptotically only.}
\label{fig3}
\end{figure}

In figure \ref{fig3} we show several cases for systems with $L_x \ne L_y$. Figure \ref{fig3}a contains the topological phases where $\chi$ reaches integer values for strong U. As in the previous cases for the square sample systems, the fillings $f=\frac{5}{4 \times 3},\frac{7}{5 \times 3},\frac{6}{5 \times 4}$ correspond to DWO with $\chi=N$, while $f=\frac{10}{5 \times 3}$ corresponds to structures with four clusters giving $\chi=4$. As shown in figure \ref{fig2} the four clusters contain no cycles and consist of one,four and seven particles distributed in various configurations inside the Hubbard lattice. Figure \ref{fig3}b contains fillings where integer values of $\chi$ are approached only asymptotically for $U \rightarrow \infty$.

We note that an alternative definition of the Euler characteristic (Eq. \ref{euler}), in terms of the numbers of clusters $C_0$ and the number of cycles $C_1$ in the grid graph formed by the particles, is given by $\chi=C_0-C_1$, known as the Euler-Poincare formula. This can be proven easily by taking account of the Euler formula for planar graphs which gives $\chi=-C_1+1$\cite{excited}, for each individual cluster and then summing this value over all the clusters to get total Euler for one Fock state.

A mechanism based on the energetics of the system
can be revealed by enumerating the sites of the Hubbard
lattice based on their coordinates x,y, which can take even or odd values. Due to the second nearest-neighbor hopping $H_t$ in Eq. 1, the Hamiltonian matrix of the system splits in diagonal blocks with each one corresponding to occupied sites in the lattice that have the same type of x,y coordinates. For example two blocks are formed when both x,y are of the same type even or odd. This block diagonal form of the Hamiltonian matrix allows us to diagonalize it more efficiently for large systems. The block containing
the lowest eigenvalue of the matrix, gives the ground state of the system. The interplay between the diagonal blocks in the Hamiltonian is partly responsible for the formation of the quantum phases with integer Euler containing either DWO or CO, that we have shown.

In addition we note that the inclusion of a small nearest-neighbor hopping in the Hamiltonian Eq. \ref{eq1} does not affect the quantum phases characterized by the integer values of the Euler characteristic.

To conclude this section, we have shown that the ground state of a 2D Hubbard model with a nearest-neighbor interaction and a second nearest hopping, forms quantum phases for strong interaction strength, that are characterized by an integer Euler number, describing the graph structures formed by the particles.
The graph structures persist from the case when the hopping term is absent and the energy of the system is fully determined by the number of bonds formed by adjacent particles in the lattice, due to the interaction term.
For half-filling or below ($f \le \frac{1}{2}$) the particles self-organize into states with density-wave-order (DWO), where nearest-neighboring sites in the Hubbard lattice cannot be simultaneously occupied. This results in Euler number equal to the number of particles $\chi=N$, since there are no edges(bonds) in the graph structures formed by the particles. Above half-filling ($f>\frac{1}{2}$) the appearance of quantum phases with integer Euler depends on the details in the structures formed by the particles as they condense into clusters. For example, we have found that in the absence of closed vertex lines(cycles) in the particle structures, the Euler is equal to the number of clusters. This clustering-order (CO) can be thought as a superfluid phase consisting of line-like clusters(strings) with no cycles(loops).  

\section{Excited States}
\begin{figure}
\begin{center}
\includegraphics[width=0.9\columnwidth,clip=true]{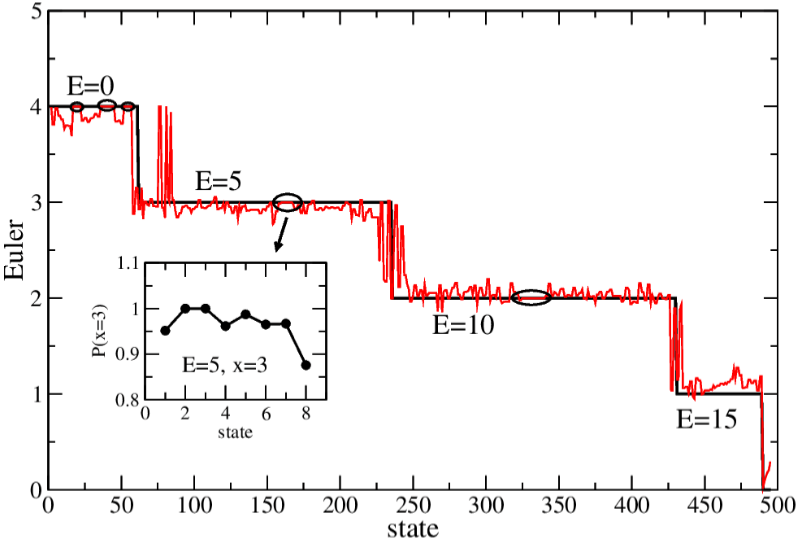}
\end{center}
\caption{Main figure: The Euler number $\chi$ versus the state index, for the whole energy spectrum of a system with filling $f=\frac{4}{4\times3}$ and interaction strength $U=5$. The black curve represents the case where the hopping term is absent ($t=0$), when the energies of the system are fully determined by the interaction term between the particles, resulting in plateaus with various integer values of $\chi$. When the hopping is added ($t=1$), the corresponding $\chi$ represented by the red curve, fluctuates around the black curve. The encircled areas contain degenerate states, where the same $\chi$ is retained for both the $t=0$ and $t=1$ cases, when the black and red curves overlap. Inset: The probability contribution of Fock states with $\chi=3$ in the superposition that forms each of the degenerate states in the encircled area indicated by the arrow.}
\label{fig4}
\end{figure}

In this section we examine the self-organization
of the interacting particles at excited states away from the ground state. In figure \ref{fig4} we plot the Euler $\chi$ versus each excited state (red curve), for a system with four particles at filling $f=\frac{4}{4\times3}$, with interaction strength $U=5$ and hopping $t=1$ in Eq. \ref{eq1}.
The black curve corresponds to $t=0$, when the energy of the system is purely determined by the interaction term in Eq. \ref{eq1}. The number of edges(bonds) $M$ between nearest-neighboring(adjacent) particles in the Hubbard lattice determines the integer energies of the system, via $M=\frac{E}{U}$. The corresponding Euler numbers are given by $\chi=N-\frac{E}{U}$ represented by the plateaus in the black curve in figure \ref{fig4}. 
The corresponding many-body states, for example at the plateau with $\chi=3$, consist of particle structures (Fock states) with three clusters, containing an edge with two adjacent particles and two free particles, arranged in various positions inside the Hubbard lattice.

\begin{figure}
\begin{center}
\includegraphics[width=0.9\columnwidth,clip=true]{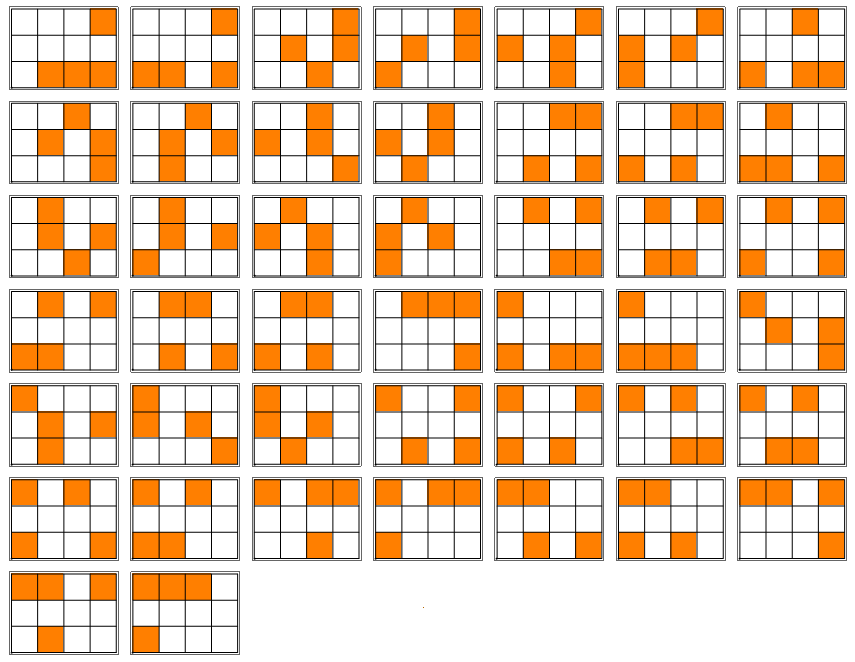}
\end{center}
\caption{The wavefunction particle structures for one of the degenerate states at filling $f=\frac{4}{4\times3}$ with $U=5$ and $t=1$, at energy $E=5$ with Euler $\chi=3$. The particle structures for each Fock state contain two to four clusters arranged in various positions inside the Hubbard lattice. The clusters are either single particles$(M=0)$ or lines of two$(M=1)$ and three adjacent particles$(M=2)$. The number of clusters inside each Fock state determines its individual Euler number, ranging from $\chi=2$ to $\chi=4$. The average Euler number over all the Fock states, given by Eq. \ref{eq_13}, retains its value $\chi=3$ from the t=0 case. }
\label{fock_test1}
\end{figure}

When the hopping term is added in Eq. \ref{eq1} ($t=1$), the result is represented by the red curve in figure \ref{fig4}. The system forms various states with values of $\chi$ that fluctuate around the plateaus in the black curve for $t=0$. The most interesting effect is encountered inside the encircled areas, which contain states that preserve the same $\chi$ for both $t=0$ and $t=1$. These states are degenerate at integer energies $E=0,5,10$ corresponding to integer Euler numbers $\chi=4,3,2$. As for the ground state, the respective many-body wavefunctions are linear superpositions of various Fock states with different particle structures. The most remarkable effect that we have found occurs inside the encircled area at the plateau with $\chi=3$ in figure \ref{fig4}. The corresponding states are linear superpositions of Fock states with different individual Euler numbers $\chi$, not necessarily equal to three. In order to demonstrate this effect, in the inset of figure \ref{fig4}, we plot the total contribution of the probability amplitudes for the Fock states with $\chi=3$ in the superposition, $P(\chi=3)$, for each one of the degenerate states inside the encircled area. For any value of the $\chi$, we have
\begin{equation}
P(\chi(t=0))=\sum_{i=Fock} |\Psi_{i}(\chi(t=0))|^2,
\label{eq_15}
\end{equation}
where $\chi(t=0)$ is the value of $\chi$ when $t=0$ in Eq. \ref{eq1} and $\Psi_{i}(\chi(t=0))$ is the corresponding probability amplitude for the Fock state i. The sum runs over Fock states that have the Euler value $\chi(t=0)$. We note that $P(\chi(t \ne 0))=1-P(\chi(t=0))$. The result of Eq. \ref{eq_15}, is plotted in the inset of figure \ref{fig4}, showing that some of the degenerate states in the encircled area with $E=5$, contain Fock states with $\chi \neq 3$. We remark that due to the degeneracy there are many possible solutions for the eigenvectors at this energy, which can be related with a unitary transformation. The eigenvectors contain the amplitudes of each Fock state in the superposition that represents each one of the degenerate states, and depend in principle on the method used to diagonalize the Hamiltonian of the system. We have verified that the Euler number of the degenerate states remains the same for different methods of diagonalization. Our result shows that the probability amplitudes in the superposition of the Fock states which forms the respective wavefunction of each degenerate state, are self-tuned in a way that retains the average Euler value $\chi = 3$ from the $t=0$ case. In this sense, the system retains the information of the self-organization of the particles from when the hopping is absent and the energy of the system is determined only by the interaction term in Eq. \ref{eq1}. This information comes from the energetical(bond) ordering of the particles, according to the number of edges(bonds) formed by adjacent particles in the Hubbard lattice. The particle structures for one of the degenerate states
at the plateau with $\chi=3$ are shown in figure \ref{fock_test1}. The individual Euler for each Fock state takes the values $\chi=2,3,4$, with the corresponding clusters arranged in various positions inside the Hubbard lattice. 

For the other encircled areas in figure \ref{fig4} at energies $E=0,10$ the self-organization information is still retained from the $t=0$ case, but the corresponding wavefunctions are linear superpositions of Fock states with individual Euler $\chi=4$ for E=0 and $\chi=2$ for E=10, as for the ground state analyzed in the previous section. 

\begin{figure}
\begin{center}
\includegraphics[width=0.9\columnwidth,clip=true]{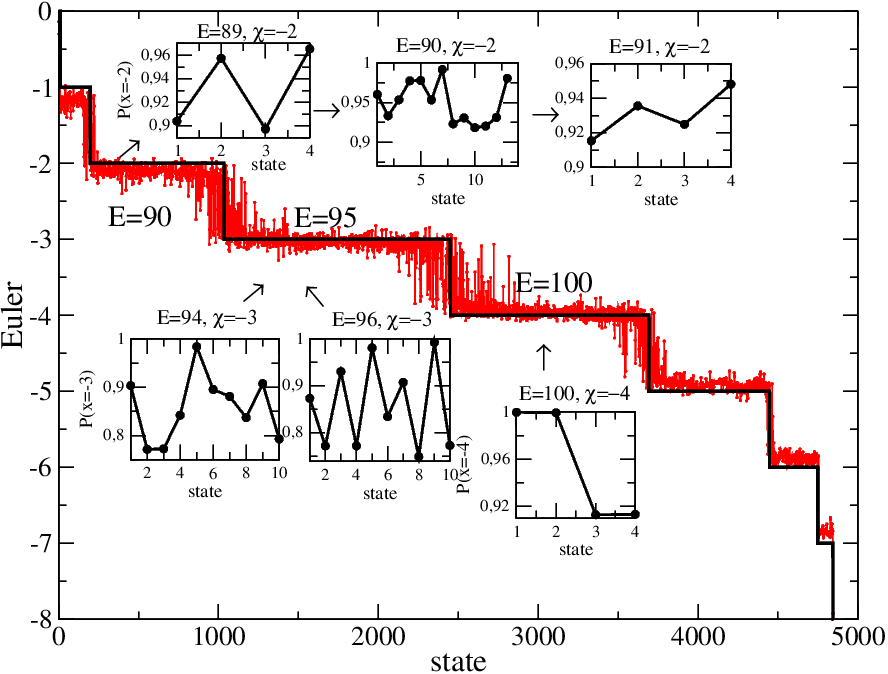}
\end{center}
\caption{Main figure: The Euler number $\chi$ versus the state index for the energy spectrum of a system with filling $f=\frac{16}{5\times4}$ and interaction strength $U=5$. The result for hopping $t=0$ is represented by the black curve which contains plateaus with integer values of $\chi$ and corresponding integer energies $E$, whose values are determined by the number of edges in the graph structures formed by the particles. For $t=1$ the $\chi$ represented by the red curve fluctuates around the black curve. The arrows indicate areas in the energy spectrum  where the same $\chi$ is retained for both $t=0$ and $t=1$, although the corresponding integer energies might differ between the two cases. Insets: The probability contribution of Fock states with $\chi(t=0)$ in the superposition, that gives each of the degenerate states in the areas indicated by the arrows, calculated via Eq. \ref{eq_15}.}
\label{fig5}
\end{figure}
\begin{figure}
\begin{center}
\includegraphics[width=0.9\columnwidth,clip=true]{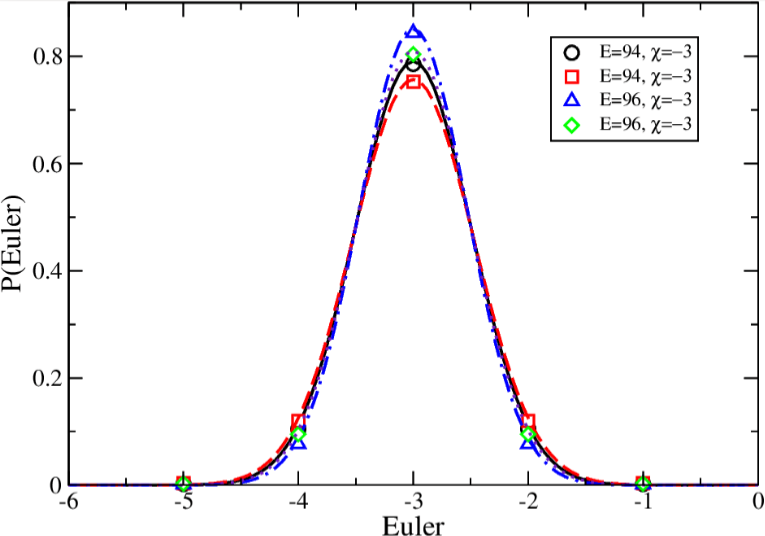}
\end{center}
\caption{The probability distribution of the individual Euler numbers for the Fock states whose superposition gives the wavefunction of the system at filling  $f=\frac{16}{5\times4}$, with $U=5$ and $t=1$. We show energies $E=94,96$ with average Euler $\chi=-3$, which corresponds to energy $E=95$ in the $t=0$ limit.}
\label{fig7}
\end{figure}
Another example for a dense system at filling $f=\frac{16}{5\times4}$ is shown in figure \ref{fig5} for $U=5$ and $t=1$. This system has four holes(empty sites) that act similarly to the four particles for the $f=\frac{4}{4\times3}$ case analyzed above. The Euler for $t=0$ takes negative values, since there are a lot of cycles (closed vertex lines) in the particle structures, due to the limited free space in the Hubbard lattice. Negative integer values like $\chi=-4,-3,-2$ are reached, represented by the plateaus in the black curve. The arrows in figure \ref{fig5} indicate places where the Euler number for $t=1$(red curve) is preserved from the $t=0$ case, although the integer energy of the corresponding states might differ from $t=0$. The contribution of the probability amplitudes for the individual Fock states with $\chi=-4,-3,-2$, calculated by Eq. \ref{eq_15} is shown in the insets, for the places in the energy spectrum indicated by the arrows.
 
Another effect that we have found is that the probability $P(\chi)$ for each individual $\chi$ in the superposition of the Fock states, spreads around the $\chi(t=0)$ value normally, via a Gaussian distribution. A few examples are shown in figure \ref{fig7} for $E=94,96$ and $\chi=-3$. From this perspective, we can see the second nearest neighbor hopping $t$ acts as a perturbation causing the Euler number to spread normally around the value it has at the $t=0$ limit.
 
The effects described for the two fillings studied above
are general and are encountered at various other fillings
that we have studied.

In conclusion, we have found that at certain areas of the energy spectrum, many-body states emerge, whose structural properties are determined by the energetical self-organization of the particles when the hopping term is absent and the properties of the system are determined by the number of edges in the graph structures formed by the particles.

\section{Summary and Conlusions}
We have studied the self-organization of a few strongly interacting particles in 2D Hubbard models with a nearest-neighbor interaction term and a second nearest-neighbor hopping. We have shown the emergence of quantum phases for strong interactions, characterized by an integer value of the Euler number describing the graph structures formed by the interacting particles as they self-organize at different fillings and energies. At the ground state the system forms quantum phases containing either density-wave-order (DWO) for low fillings, where the particles cannot occupy adjacent sites in the Hubbard lattice, or clustering-order(CO) for large fillings, where the particles condense into clusters with various interesting properties related for example to topology. The structural properties of these phases are determined by the number of edges in the graph structures formed by the particles, i.e. on the energetical self-organization of the particles, when the hopping term is absent and the energy of the system is fully determined by the interaction term only. More surprisingly for the excited states we have found various quantum phases that retain the value of the Euler characteristic from the case when the hopping is absent. The corresponding wavefunction amplitudes of the Fock states in the superpositions that describe the quantum phases,
are self-tuned in a way that retains the value of the Euler number from when the hopping term is absent. In conclusion we have demonstrated various quantum phases with novel structural properties emerging in simple Hubbard models containing a few strongly interacting particles. These phases could be potentially realized in cold-atom experiments.

 \section*{Acknowledgements}
We acknowledge resources, infrastructure and financial support provided by the Project HPC-EUROPA3 (INFRAIA-2016-1-730897), funded by the EC Research Innovation Action under the H2020 Programme, GRNET and the ARIS-GRNET computing network, along with the Physics Department at the University of Ioannina in Greece.

\section*{Data Availability Statement}
This manuscript has no associated data or the data will not be deposited. All the numerical data from our calculations are displayed/plotted inside the figures.

\section*{Author Contribution Statement}
Both authors I.K. and I.A. contributed equally to the design and implementation of the research, to the analysis of the results and to the writing of the manuscript.


\end{document}